# Predicting cognitive load in immersive driving scenarios with a hybrid CNN-RNN model

Mehshan Ahmed Khan[1][0000-0002-7380-8532], Houshyar Asadi[1] [0000-0002-3620-8693], Mohammad Reza Chalak Qazani [2] [0000-0003-1839-029X], Adetokunbo Arogbonlo[1] [0000-0002-4524-8267], Saeid Nahavandi[3] [0000-0002-0360-5270] and Chee Peng Lim[1] [0000-0002-4524-8267]

[1] Institute for Intelligent Systems Research and Innovation, Deakin University, VIC, Australia
[2] Faculty of Computing and Information Technology (FoCIT), Sohar University, Sohar, Oman
[3] Swinburne University of Technology, Hawthorn, Victoria, 3122, Australia
{mehshan.khan, houshyar.asadi, a.arogbonlo, chee.lim}@deakin.edu.au, m.r.chalakqazani@gmail.com, snahavandi@swin.edu.au

**Abstract.** One debatable issue in traffic safety research is that cognitive load from secondary tasks reduces primary task performance, such as driving. Although physiological signals have been extensively used in driving-related research to assess cognitive load, only a few studies have specifically focused on high cognitive load scenarios. Most existing studies tend to examine moderate or low levels of cognitive load In this study, we adopted an auditory version of the n-back task of three levels as a cognitively loading secondary task while driving in a driving simulator. During the simultaneous execution of driving and the n-back task, we recorded fNIRS, eye-tracking, and driving behavior data to predict cognitive load at three different levels. To the best of our knowledge, this combination of data sources has never been used before. Unlike most previous studies that utilize binary classification of cognitive load and driving in conditions without traffic, our study involved three levels of cognitive load, with drivers operating in normal traffic conditions under low visibility, specifically during nighttime and rainy weather. We proposed a neural network combining a 1D Convolutional Neural Network and a Recurrent Neural Network to predict cognitive load. Our experimental results demonstrate that the proposed model, with fewer parameters, increases accuracy from 99.82% to 99.99% using physiological data, and from 87.26% to 92.02% using driving behavior data alone. This significant improvement highlights the effectiveness of our CNN-RNN in accurately predicting cognitive load during driving under challenging conditions.

**Keywords:** N-back task, Cognitive load, Machine learning, Deep learning, CNN-RNN model.

## 1 Introduction

Today's vehicle systems are more advanced, faster, and safer than ever before, progressing towards full autonomy [1]. Despite these advancements, cognitive load remains a significant factor in traffic crashes during manual driving, as reported by various surveys [2]. Until fully autonomous driving becomes a reality, drivers must continue to monitor the vehicle and take control in critical situations. Engaging in secondary tasks, such as conversing with a passenger while driving, can lead to inattention to the primary task of driving [3]. This inattention can manifest in different forms: visual, manual, or cognitive [4]. While visual and manual distractions are crucial components of driver monitoring, our focus is on detecting phases with high cognitive load.

High cognitive load is particularly important because, in dual- or multitask situations, a driver's attention is a finite resource, and cognitive limitations are quickly reached. The research community uses various terms to describe a subject's cognitive state, such as cognitive workload [5], mental workload [6], or task load [7]. In this study, we use the term Working Memory (WM) to refer to the storage of conscious information, which is limited by the amount one can hold and process. Cognitive Load (CL) is used as a measure to define the amount of load placed on the WM during a task [8, 9].

Previous studies have explored the prediction of driver fatigue using state-of-the-art Machine Learning (ML) [10] and Deep Learning (DL) [11] techniques, leveraging recent advancements in computer vision, pattern recognition, image segmentation, and signal processing [12, 13]. These techniques have significantly improved the ability to detect and predict driver fatigue by analyzing a wide range of physiological signals and behavioral indicators [14]. Physiological signals such as electro-cardiogram [15], functional near-infrared spectroscopy (fNIRS) [16], eye-tracking [17], and electrodermal activity [18] are commonly used in these studies. These studies were predominantly conducted in driving simulators [19, 20], which provide a controlled environment to simulate various driving conditions and monitor driver responses [21]. To induce different levels of cognitive workload, participants, typically ranging from 10 to 30 individuals were asked to engage in prolonged manual driving sessions. These sessions were usually set on a highway in a monotonous environment devoid of traffic, ensuring a constant and reasonable speed. During these sessions, participants simultaneously performed the n-back task, a well-known method for imposing cognitive load [22], to create varying levels of mental workload. This involved applying a range of machine learning and deep learning techniques to the collected data. The primary goal was to develop models capable of accurately predicting the driver’s cognitive state based on the physiological signals and driving behavior. By leveraging ML and DL techniques, researchers have been able to identify patterns and correlations within the data that are indicative of driver fatigue or increased cognitive load. These predictive models are crucial for developing advanced driver assistance systems (ADAS) that can monitor a driver’s state in real-time and provide timely interventions to prevent accidents.

Despite the progress made in predicting driver fatigue and cognitive load using machine learning techniques, several challenges remain. One significant challenge is the introduction of additional variables in real-world driving conditions. Many existing studies have analyzed drivers' cognitive load without considering realistic driving conditions such as nighttime driving or adverse weather conditions like rain. These factors can significantly impact cognitive load and driving behavior, yet they are often neglected in research. Most previous studies have focused on binary classification tasks, distinguishing between different levels of workload or fatigue. However, these approaches do not fully capture the complexities of real-world driving scenarios. To address this gap, the aim of this study is to analyze drivers' cognitive load using the auditory version of the n-back task in low visibility conditions, specifically nighttime and rainy weather. These conditions have often been overlooked in previous research, despite their importance in understanding cognitive load in realistic settings. This study builds on previous work Khan et al. [23], where they applied a Convolution Long Short-Term Memory (Conv-LSTM) model with two convolutional layers and two LSTM layers to predict cognitive load. This approach allowed for a more comprehensive analysis of how cognitive load evolves over time and in different spatial contexts. Additionally, previous research often relied on simplified or artificial driving scenarios that did not fully represent the complexities of real-world driving. In contrast, Khan et al. [23] used a dataset collected from participants in more realistic settings, providing a richer and more accurate representation of the driving experience and making their findings more applicable to real-world conditions. The dataset included fNIRS, eye-tracking, and driving behavior data recorded while participants drove in nighttime and rainy conditions in a driving simulator that provides a realistic driving experience. In the present study, we will apply a Convolutional Recurrent Neural Network (CNN-RNN) model with two convolutional layers and RNN layer to the same dataset. In the present study, we will apply a Convolutional Recurrent Neural Network (CNN-RNN) model with two convolutional layers and a recurrent neural network (RNN) layer to the same dataset. This approach is designed to leverage the strengths of both convolutional and recurrent neural networks. Additionally, we have employed the Extra Trees classification method to select the optimal number of features from fNIRS signals. To evaluate the effectiveness of our proposed model, we will compare its performance with other feature selection methods. These methods might include traditional techniques such as Principal Component Analysis (PCA), ANOVA, and other advanced machine learning-based feature selection algorithms. By conducting this comparative analysis, we aim to determine whether the Extra Trees classification method provides a superior selection of features that enhance the model's performance in predicting cognitive load. Moreover, we will also compare the results of our current study with our previous study, which utilized the Analysis of Variance (ANOVA) feature selection method. The previous study incorporated a large number of model parameters, aiming to identify and use the most relevant features for predicting cognitive load. This analysis will help us determine if the Extra Trees method provides a better selection of features that lead to improved model performance, or if the ANOVA method with a larger number of model parameters yields superior results. The goal of this comparison is to identify the most effective approach for feature selection in the context of predicting cognitive load from fNIRS signals. By doing so, we aim to contribute to the optimization of cognitive load assessment models, ultimately enhancing their applicability and reliability in real-world driving conditions. This comprehensive evaluation will highlight the trade-offs between different feature selection methods and model complexities, guiding future research and development in this field.

The subsequent sections of this paper explore various aspects of the conducted research. In Section 2, we provide a detailed account of the equipment used for data collection and the experimental procedures followed. Section 3 presents a thorough overview of the proposed methodology and its characteristics. In Section 4, we analyze cognitive load using deep learning (DL) methods. Finally, Section 5 summarizes the findings and highlights the study's significance.

## 2 Materials and methods

The dataset used in this study is based on the dataset of the study by Khan et al. [23]. The experiment was conducted using a driving simulator, namely the "Next Level Racing Motion Platform V3". Driving simulators are valuable tools designed to provide a realistic driving experience in a virtual environment [24-26], with primary goals including enhancing training programs, improving road safety, and enabling researchers to analyze driver behavior [27-30]. To simulate nighttime and stormy weather conditions, the Euro Truck Simulator 2 (ETS2) software was employed. At the beginning of the experiment, the car was parked in the emergency lane of a freeway with a speed limit of 100 km/hr, as shown in **Fig. 1**a. Participants were instructed to drive on the freeway without taking any exits, as depicted in **Fig. 1**b. To induce three different levels of mental workload—0-back, 1-back, and 2-back—we used the auditory version of the n-back task. Psychopy [31] was utilized to implement the n-back task and record responses. The design ensured that participants heard both the digit sounds and their surrounding environment simultaneously, adding an additional layer of cognitive complexity during the driving simulation.

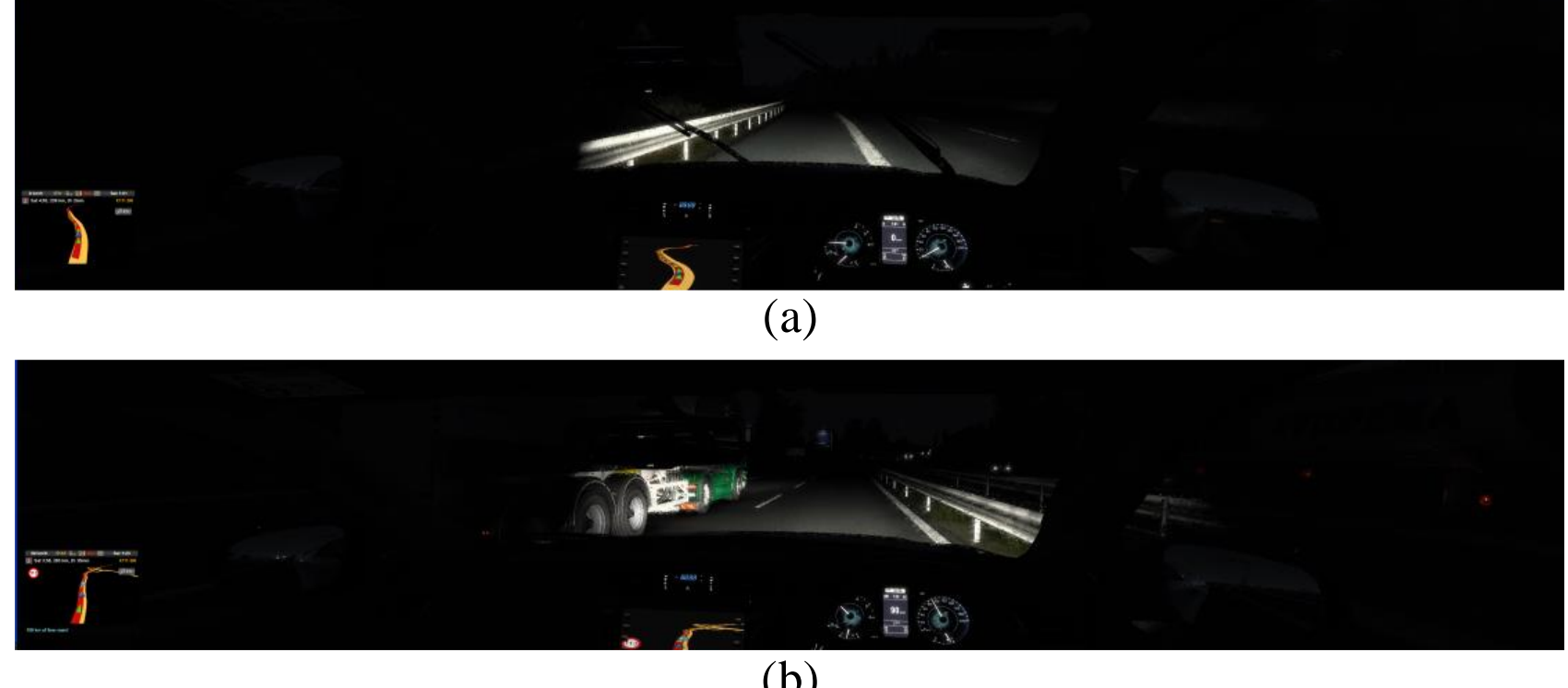

(a)

(b)

**Fig. 1.** Experimental setup for the driving simulation. (a) Car parked in the emergency lane of a freeway with a speed limit of 100 km/hr at the start of the experiment. (b) Freeway driving environment where participants drove without taking exits, under simulated nighttime and stormy weather conditions.

We collected driving data alongside fNIRS and eye-tracking data to predict cognitive load at different levels as shown in **Fig. 2** For the fNIRS signals, we used the OBELAB device, and for eye-tracking data, we used the Pupil Core glasses. Additionally, we recorded car speed, angular velocity (X, Y, Z), linear acceleration (X, Y, Z), steering wheel angle, throttle response, and braking response using the ETS2 Software Development Kit (SDK). The study involved a cohort of 10 healthy adults, consisting of 9 males and 1 female, who were selected through a combination of campus and online recruitment strategies.

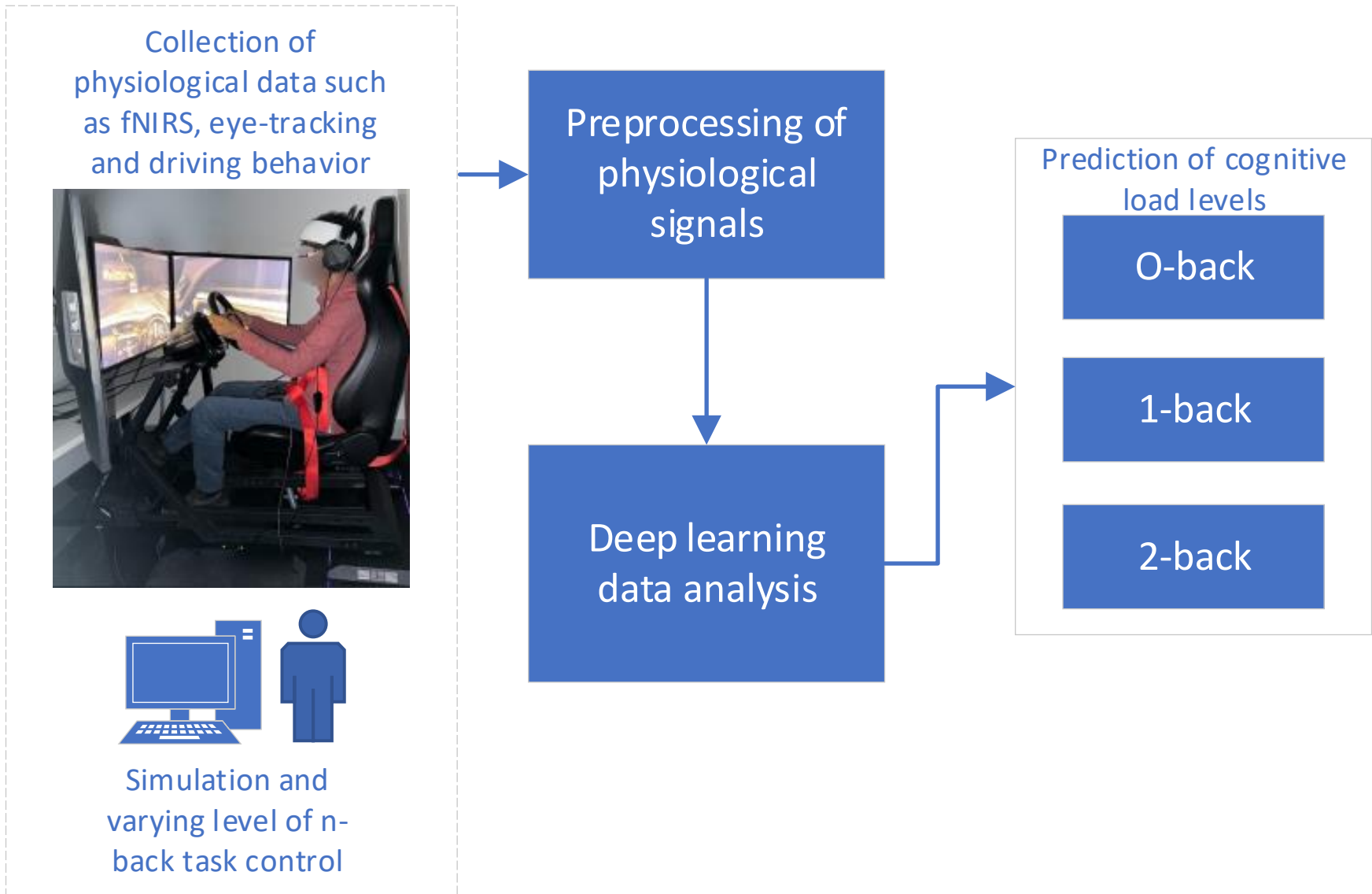

**Fig. 2.** Prediction of cognitive load in a driving simulation environment.

This comprehensive data collection approach allowed us to capture a wide range of variables relevant to driving performance and cognitive load, providing a robust dataset for subsequent analysis and model training. The inclusion of real-time physiological data, combined with detailed driving behavior metrics, enhances the potential for accurate and reliable predictions of cognitive load under various driving conditions.

## 3 Proposed CNN-RNN based architecture

Recent advancements in the field of cognitive load classification have seen the integration of CNNs and Recurrent Neural Networks (RNNs) to analyze various physiological signals [32, 33]. These models have been employed for the classification of cognitive load using EEG, ECG, heart rate, and fNIRS signals [34, 35]. The CNNs are particularly adept at recognizing spatial patterns in data, which is crucial for interpreting complex physiological signals. On the other hand, RNNs excel in processing sequential data, making them suitable for tasks that require an understanding of temporal dependencies [36, 37]. This approach offers significant advantages over traditional models, as it combines the strengths of both CNNs and RNNs. CNNs can extract intricate features from raw data, such as fNIRS signals, while RNNs can capture the temporal evolution of cognitive load. This combination is especially effective for complex tasks like cognitive load analysis from multiple data sources, including fNIRS signals, eye-tracking data, and driving behavior metrics. In our study, we utilized this CNN and RNN-based model to predict cognitive load in a simulated driving environment. By integrating fNIRS signals, eye-tracking data, and driving behavior metrics, our approach aims to provide a comprehensive assessment of cognitive load. This integrated model allows for the capture of both spatial features and temporal dynamics, leading to more accurate and reliable predictions of cognitive load levels in complex and demanding driving scenarios.

### 3.1 Proposed CNN and RNN based model

The model architecture incorporates both CNN [12] and RNN [38] layers, capitalizing on their respective strengths. CNN layers excel at extracting spatial features from input data, enabling robust feature representation through convolutional operations. Meanwhile, RNN layers specialize in capturing temporal dependencies within sequential data, facilitating the modeling of long-range dependencies and dynamic patterns over time. By adopting a 1D CNN and RNN architecture, our system efficiently harnesses the strengths of both architectures for feature training.

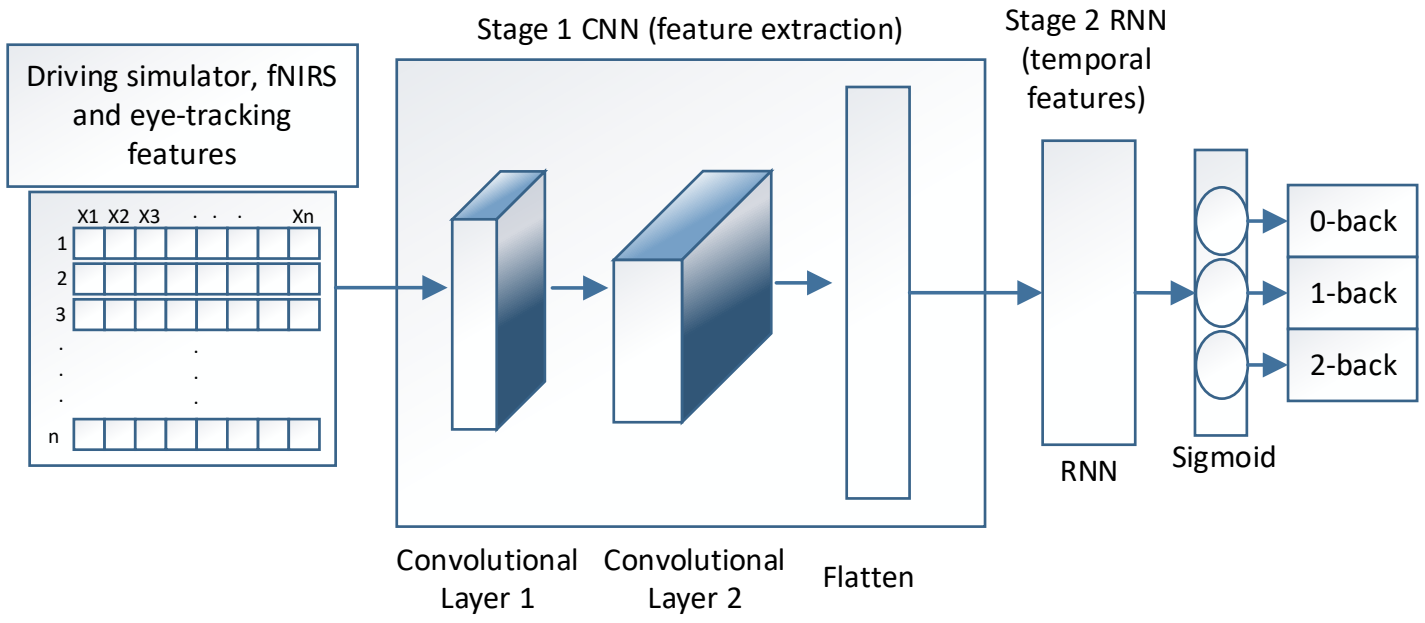

**Fig. 3.** Architecture of the Proposed CNN-RNN Model

In the workflow of our CNN–RNN model, depicted in **Fig. 3**, the dataset undergoes preprocessing before being divided into training and testing sets. Subsequently, data normalization is carried out to ensure uniformity and facilitate effective training. The normalized data is then inputted into the model, starting with two CNN layers. These CNN layers have kernel sizes of 3 and output channels of 16 and 32, respectively, followed by rectified linear unit (ReLU) activation functions to capture spatial features from the input data. Next, the RNN layer, with an input size of 32, a hidden size of 64, and two layers, captures temporal dependencies in sequential data. Configured with a batch-first arrangement, the model flattens the RNN output and passes it through one fully connected layer of size 64, followed by a sigmoid activation function to interpret the final output. This architecture allows our model to effectively learn and predict complex patterns within the data.

# 4 Results and discussion

To begin the assessment of cognitive load, both physiological sensor data and driving data from participants underwent comprehensive preprocessing to ensure uniformity and suitability for analysis. The standardized scalar function was applied to both the driving simulator data and physiological measures, normalizing the data to a mean of zero and a standard deviation of one, ensuring all features were on a standardized scale. Specific features relevant to cognitive load assessment were extracted, including steering wheel angle, throttle response, braking response, car speed, angular velocities, and linear accelerations from driving data, as well as fNIRS-derived oxygenation levels and eye-tracking metrics from physiological data. These features were temporally aligned to ensure synchronicity, accurately associating cognitive load with specific driving events and physiological responses. To optimize model performance, we compared various dimensionality reduction techniques, such as Principal Component Analysis (PCA), ANOVA, variance threshold, and Extra Tree Classifier, to manage the complexity and potential redundancy in the high-dimensional feature space, retaining the most informative features while simplifying the analysis. These preprocessing steps prepared the data, providing a reliable foundation for the subsequent analysis of cognitive load using the proposed Conv-RNN model, thereby improving data quality, consistency, and the accuracy of the cognitive load assessment.

## 4.1 Extra tree feature analysis

We employed the Extra Tree Classifier [39], which is a meta estimator that fits multiple randomized decision trees (also known as extra-trees) on various sub-samples of the dataset, using averaging to improve predictive accuracy and control overfitting. This classifier computes feature importance, allowing for the discarding of irrelevant features. Extra trees generate numerous individual decision trees from the entire training dataset. For the root node, the algorithm selects a split rule based on a random subset of features (k) and a partially random cut point, dividing the parent node into two random child nodes. This process continues recursively in each child node until a leaf node is reached, where no further splits occur. During the construction of the forest, the Gini importance for each feature is computed to perform feature selection. The Gini importance measures the contribution of each feature to the impurity reduction at each split in the decision trees. Features are then ordered in descending order based on their Gini importance scores.

### 4.1.1 fNIRS feature selection.

The fNIRS system records blood oxygenation and deoxygenation levels across 204 channels. However, not all of these channels are equally useful for predicting drivers' cognitive load. To identify the most relevant channels, we computed the feature importance for each fNIRS channel. Fig. 4 shows the feature importance plot of fNIRS channels, obtained using the Extra Tree Classification method, showing the relative importance of each channel in decreasing order. By analyzing the feature importance plot, we can determine which channels contribute most significantly to the prediction of cognitive load. Channels with higher importance scores are more relevant and likely to provide valuable information for our model. Conversely, channels with lower importance scores may contribute little to the prediction accuracy and can be considered for removal to streamline the dataset and reduce computational complexity. The feature importance plot is crucial for several reasons. First, it helps in understanding the spatial distribution of relevant brain activity as captured by the fNIRS channels, which can offer insights into the neural correlates of cognitive load during driving. Second, it guides the feature selection process, ensuring that our model is both efficient and effective by focusing on the most informative channels. This selective approach enhances the model's performance and reduces the risk of overfitting, particularly when dealing with high-dimensional data like fNIRS.

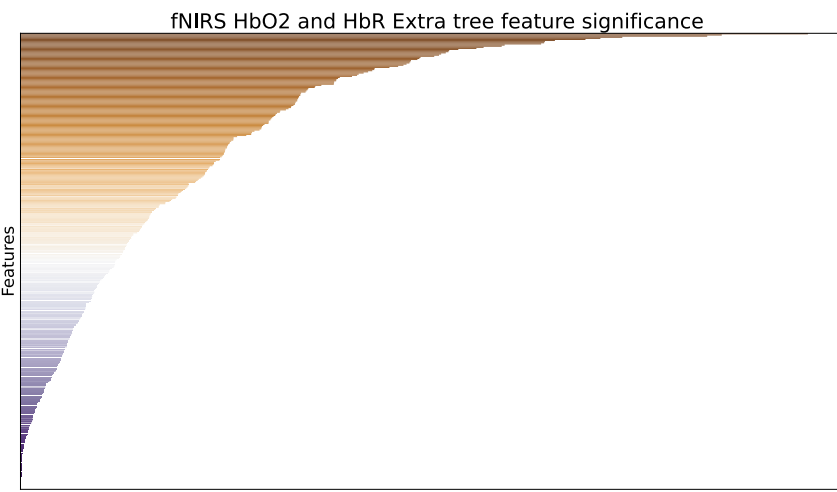

**Fig. 4.** fNIRS channel importance for cognitive load prediction.

To optimize the performance of our model, we selected the top 20 features from the fNIRS channels. **Fig. 5**a illustrates the feature importance plot of these top 20 features. Features whose names end with "O2" represent channels responsible for capturing blood oxygenation levels, while features ending with "R" represent channels responsible for capturing blood deoxygenation levels. The feature selection using the Extra Trees Classification method reveals that the importance is fairly evenly distributed among the top features. However, in a previous study where we utilized the ANOVA feature selection method, as shown in **Fig. 5**b, only the first two features had similar importance, while the remaining features showed significantly less importance. Interestingly, both feature selection methods indicate that oxygenation channels are more prevalent among the top features compared to deoxygenation channels, suggesting that oxygenation levels are more critical for predicting cognitive load.

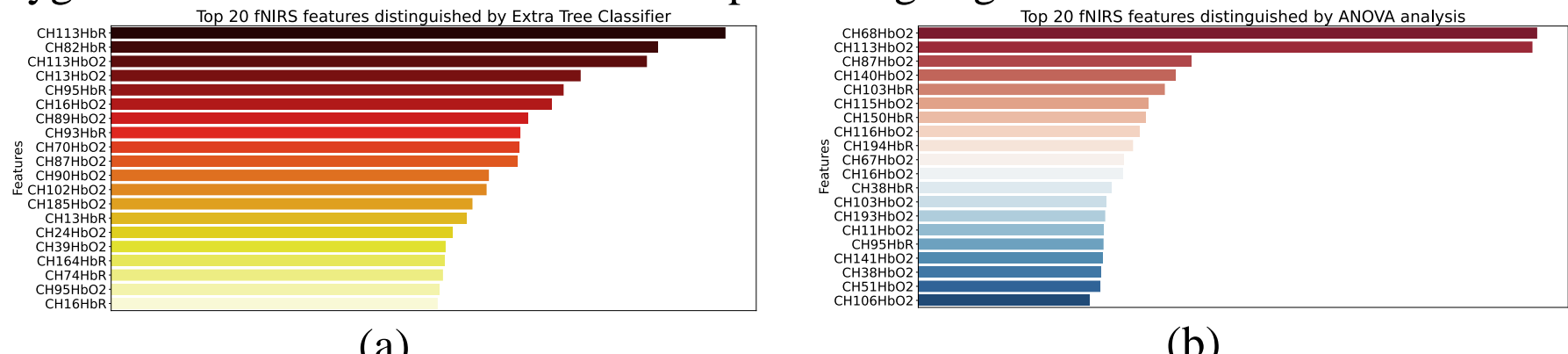

**Fig. 5.** Importance of the top 20 fNIRS features for cognitive load prediction (a) Extra trees feature selection method (b) ANOVA.

#### 4.1.2 Driving behavior features analysis.

To analyze which features from the driving simulator data contribute most significantly to predicting cognitive load, we applied the Extra Tree Classification method. As shown in Fig. 6a, the feature importance plot reveals that car speed and throttle response are the most affected by different cognitive load conditions. These features have the highest importance scores, indicating their strong influence on the model's predictive capability. In contrast, our previous study utilized the ANOVA feature selection method, which produced different insights. As shown in Fig. 6b, ANOVA analysis indicated that, following car speed, braking behavior was significantly impacted by cognitive load. This suggests that car brakes are also an important factor in predicting cognitive load, albeit to a lesser extent than speed.

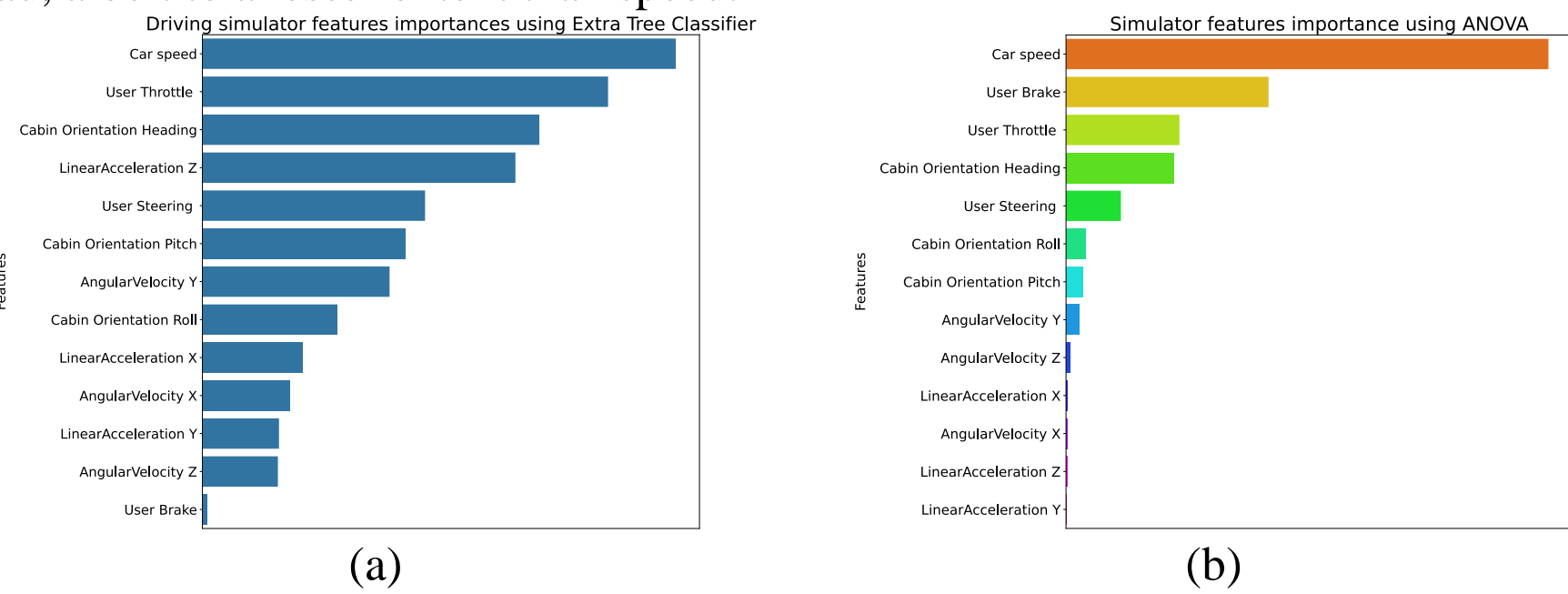

**Fig. 6.** Importance of various driving behavior features for predicting cognitive load (a) Feature importance plot using Extra Tree Classification (b) Feature importance plot using ANOVA

In the Extra Tree Classification method, the use of brakes appears as the least contributing factor in classifying cognitive load. This suggests that, while braking may be influenced by cognitive load, its impact is less pronounced compared to car speed and throttle response. The difference in findings between the two methods underscores the importance of using multiple feature selection techniques to gain a comprehensive understanding of which features are most relevant for predicting cognitive load. These differences highlight the importance of employing multiple feature selection techniques to gain a comprehensive understanding of which features are most relevant for cognitive load prediction. By combining information from both methods, we can optimize our predictive models to enhance their accuracy and robustness. This approach allows us to better understand the dynamics of driving behavior under varying cognitive loads, ultimately contributing to more effective monitoring and management strategies in both simulated and real-world driving scenarios.

#### 4.1.3 Feature independence and correlation

To combine the different physiological signals and driving behavior data effectively, we employed a down sampling approach. The driving simulator data is originally sampled at a much higher frequency compared to the fNIRS and eye-tracking signals. This discrepancy in sampling rates can pose significant challenges when attempting to integrate these diverse data streams into a cohesive dataset. To address this, we down sampled the driving simulator data to match the lower sampling frequency of the fNIRS and eye-tracking signals. Down sampling involves reducing the sampling rate of the driving simulator data, which retains the essential characteristics and patterns of the original data while discarding excess points. This process ensures that the time-series data [40] from all sources are aligned temporally, facilitating accurate synchronization.

After combining the data from different modalities, it is crucial to check feature independence to ensure that the combined features represent their unique characteristics and do not exhibit high correlation with other features. To achieve this, we performed a correlation matrix analysis. This analysis involves computing the pairwise correlation coefficients between all features, which helps in summarizing the data and understanding the relationships between features and targeted outcomes. The correlation matrix provides a comprehensive overview of how each feature relates to the others. By representing these relationships in a heat map, as depicted in **Fig. 7**, we can visually assess the degree of correlation between features.

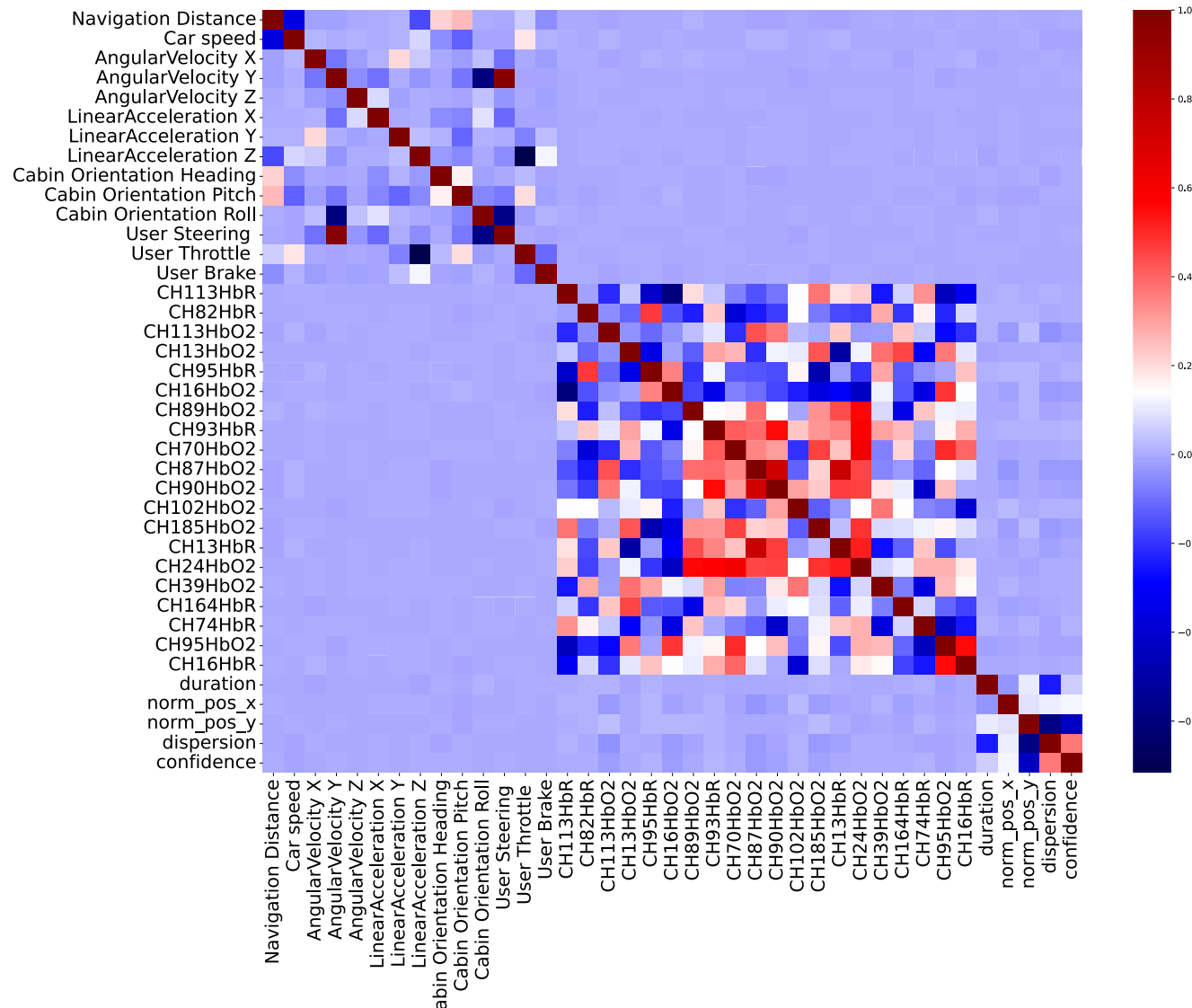

**Fig. 7.** Visualization of the correlation matrix depicting the relationships between features derived from the driving simulator, fNIRS, and eye-tracking data

In our analysis, we found that the features derived from the driving simulator data, fNIRS, and eye-tracking signals exhibit some correlations within their respective categories. For instance, certain features within the driving simulator data, such as car speed and throttle response, may be correlated due to their inherent relationship in driving behavior. Similarly, features within the fNIRS data, like various channels measuring blood oxygenation levels, may show correlations due to the physiological processes they capture. However, the critical finding from our correlation matrix analysis is that features from different sensor modalities driving simulator, fNIRS, and eye-tracking do not exhibit high correlations with each other. This independence between modalities is essential as it indicates that each type of data contributes unique information to the combined dataset, without redundancy. This lack of cross-modality correlation ensures that the integrated dataset provides a more detailed view of cognitive load, capturing diverse aspects of driver behavior and physiological responses.

### 4.2 Proposed DL based model classification results

This study employs Python, specifically utilizing the PyTorch framework, for the development and training of deep learning models. The computational setup includes an Intel Core i9-12900K CPU, 32 GB of RAM, and an NVIDIA RTX 3090 Ti GPU to manage the complex tasks of model training and evaluation.

For training our deep learning model, we utilized the Adam optimizer with a learning rate of 0.001. The Adam optimizer is particularly effective for training neural networks because it dynamically adjusts the learning rate throughout the training process, enabling it to handle the varying gradients of the loss function efficiently. This adaptability helps in accelerating the convergence of the model. Additionally, we integrated the cross-entropy loss function into our model architecture. The cross-entropy loss function is widely used for classification problems as it measures the performance of a classification model. The model's output layer employs the sigmoid activation function, which is appropriate for binary classification tasks as it squashes the output to a range between 0 and 1, thereby providing probabilistic interpretations of the model's predictions. We used a 70-30 split of the data, with 70% used for training and 30% for validation and testing. We trained the model for 1000 epochs. This extensive training duration ensures that the model can learn from the data, allowing it to capture complex patterns and improve its predictive performance.

With these parameters, we evaluated the proposed model by conducting an extensive analysis of the combined fNIRS, eye-tracking, and driving behavior data. Our approach utilized the Extra Trees classification method for feature extraction, which significantly contributed to the model's performance. The Extra Trees classification method, which involves constructing a multitude of decision trees and using their aggregated results for predictions, proved highly effective in identifying the most relevant features from the combined dataset. This method not only improves predictive accuracy but also controls overfitting by averaging the results of many trees. By employing the Extra Trees classification method, we achieved an accuracy of 99.91%. In addition to using the Extra Trees feature extraction method, we compared the model's performance with different feature extraction methods, such as Principal Component Analysis (PCA), Analysis of Variance (ANOVA), and variance threshold methods, shows in Table 1. Each of these methods has its own strengths and applications in reducing dimensionality and selecting the most informative features. The proposed model achieved higher accuracy, precision, recall, and F1-score when trained with features extracted using the Extra Trees classification method compared to other feature extraction methods.

**Table 1.** Model performance comparison across various feature extraction methods

| Feature extraction methods | Accuracy | F1-score | AUC | Precision | Recall |
|---|---|---|---|---|---|
| Variance threshold | 0.9783 | 0.9784 | 0.9928 | 0.9784 | 0.9783 |
| PCA | 0.9060 | 0.9063 | 0.9573 | 0.9070 | 0.9060 |
| ANOVA | 0.9960 | 0.9960 | 0.9980 | 0.9960 | 0.9960 |
| Extra tree features | **0.9991** | **0.9991** | **0.9999** | **0.9991** | **0.9991** |

The confusion matrices in Fig. 8a, Fig. 8b, Fig. 8c and Fig. 8d clearly demonstrate the superiority of the Extra Trees feature extraction method over PCA, ANOVA, and variance threshold methods. The Extra Trees method exhibits negligible misclassification, as shown by the near-perfect diagonal elements in its confusion matrix. In contrast, the other methods show higher misclassification rates, indicating their relative inefficiency in capturing the most informative features for accurate cognitive load prediction.

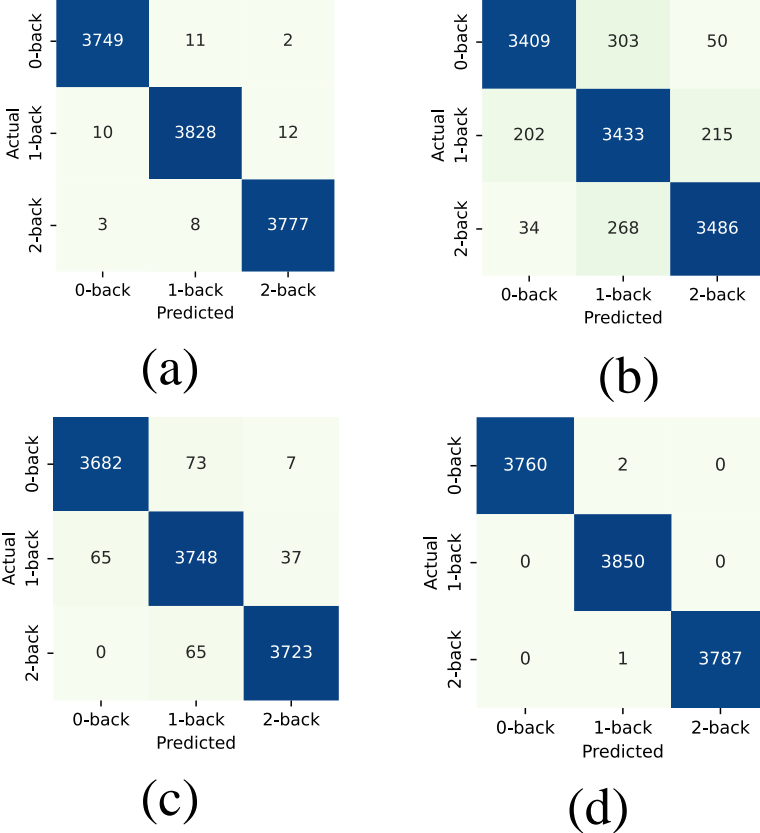

Fig. 8. Confusion matrices depicting the classification performance of the proposed model with different feature extraction techniques: (a) ANOVA, (b) PCA, (c) Variance Threshold and (d) Extra tree.

We also compared the performance of the proposed model with the model from our previous study to assess improvements and validate the efficacy of our new approach. This comparison is visualized in Fig. 9a and Fig. 9b, which highlight the performance differences between the proposed CNN-RNN model and the previous CNN-LSTM model.

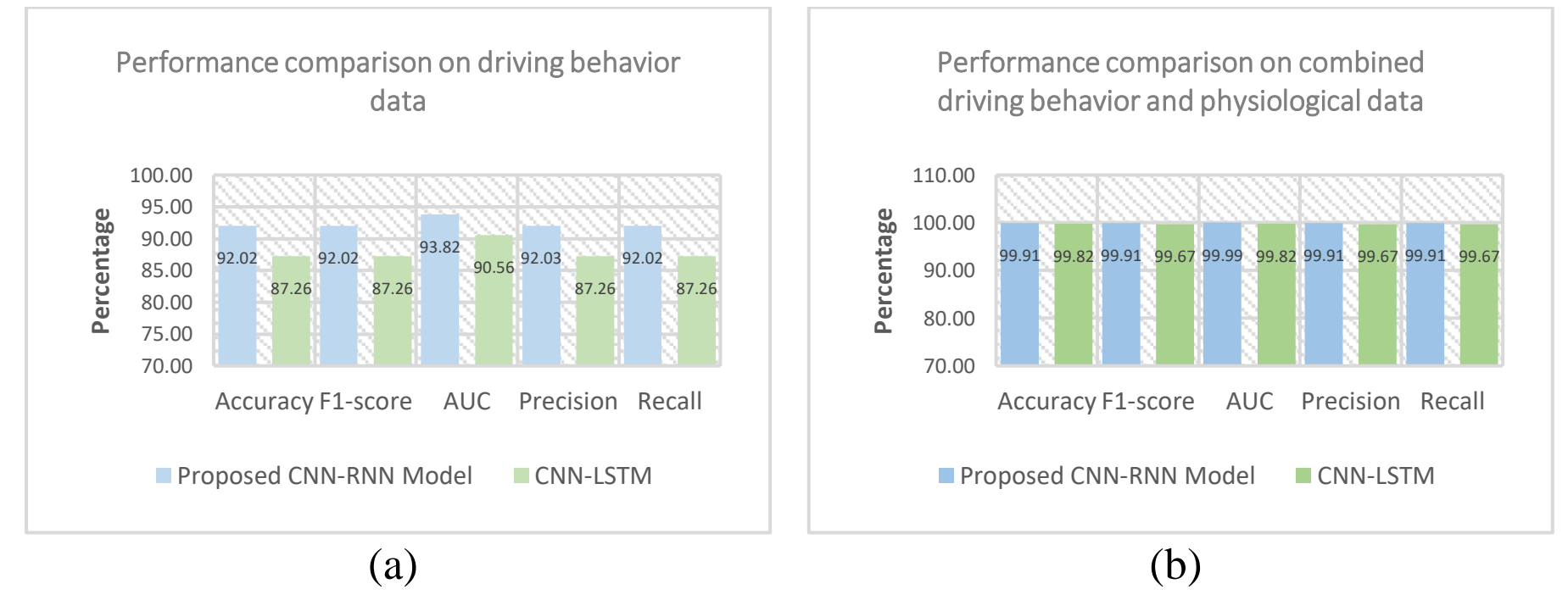

Fig. 9. Performance comparison between the proposed CNN-RNN model and the previous CNN-LSTM model using (a) combined physiological and driving behavior data (b) only driving behavior data.

Fig. 9a presents the performance evaluation of both models when trained and evaluated using the combined physiological and driving behavior data. The combined data includes inputs from fNIRS, eye-tracking, and various driving metrics. The results indicate a slight increase in performance for the proposed CNN-RNN model compared to the previous CNN-LSTM model. This enhancement, although subtle, demonstrates the proposed model's ability to better integrate and leverage the multi-modal data, leading to more accurate predictions of cognitive load. Fig. 9b shows the performance comparison when the models are trained and evaluated using only the driving behavior data. This data set includes features such as car speed, throttle response, braking, angular velocities, and linear accelerations. The proposed CNN-RNN model exhibited a significant improvement in accuracy, increasing from 87.26% to 92.02%. This substantial enhancement highlights the proposed model's superior capability in handling driving behavior data, likely due to its more effective architecture in capturing temporal dependencies and patterns within the driving metrics. These figures collectively highlight the advancements made by the proposed CNN-RNN model over the previous CNN-LSTM model. The CNN-RNN model's architecture, which combines convolutional layers for feature extraction with recurrent layers for sequence modeling, offers a powerful approach for handling complex data types. The convolutional layers are adept at capturing and extracting spatial features from various input data, such as fNIRS signals and eye-tracking metrics, by applying filters that detect local patterns and significant characteristics. Meanwhile, the recurrent layers are specialized for sequence modeling, allowing the architecture to capture temporal dependencies and dynamic patterns over time, which is crucial for accurately analyzing time-series data like driving behavior. Furthermore, incorporating the decision trees feature selection method enhances the model's robustness. Decision trees help in identifying and selecting the most relevant features from the multi-modal dataset, thereby reducing dimensionality and improving the efficiency and accuracy of the model. By focusing on the most important features, the model can better generalize from the training data and make more precise predictions.

# 5 Conclusion

Cognitive load in challenging environments, particularly in rainy weather and low visibility conditions, and under high levels of cognitive load, has been rarely addressed in the literature. These conditions are crucial to study as they significantly impact a driver’s ability to process information, make decisions, and respond to dynamic road situations. In this study, we addressed the challenging issue of predicting cognitive load in cognitively demanding driving conditions. Specifically, we focused on predicting high levels of cognitive load by integrating an auditory n-back task as a secondary cognitive task. We collected comprehensive data, including fNIRS signals, eye-tracking metrics, and driving behavior indicators including car speed, angular velocity (X, Y, Z), linear acceleration (X, Y, Z), steering wheel angle, throttle response, and braking response, under immersive driving conditions characterized by low visibility, such as nighttime and rainy weather using a driving simulator. Our approach leveraged a neural network combining 1D CNNs and RNNs to effectively analyze both spatial and temporal features of the multi-modal data. The proposed CNN-RNN model demonstrated significant improvements in accuracy compared to previous methods. By reducing the number of parameters, our model achieved an accuracy increase from 99.82% to 99.91% using physiological data, and from 87.26% to 92.02% using driving behavior data alone. These results highlights the robustness and efficiency of our architecture in capturing complex patterns and dependencies within the data, facilitating more precise predictions of cognitive load levels.

In the study [23], the ANOVA feature selection method is utilised. However, the model did not perform better with the ANOVA feature selection method compared to other techniques. In contrast, the decision tree feature selection method yielded better results, even with a model having fewer parameters. The decision tree approach enhanced the model's ability to identify and prioritize the most relevant features, significantly improving overall model performance. Incorporating the decision trees feature selection method not only optimized the feature selection process but also enhanced the model's robustness. This method allowed for a more effective training process by reducing dimensionality and concentrating on the most impactful features. Our findings suggest that the combination of CNNs, RNNs, and decision tree-based feature selection provides a powerful framework for analyzing cognitive load in driving, with potential applications in developing advanced driver assistance systems and enhancing traffic safety.

Future research should focus on enhancing the prediction of cognitive load using vehicle dynamics alone, while incorporating experiments within a higher degree-of-freedom motion simulator equipped with advanced motion cueing algorithms [41-44]. Additionally, it is crucial to validate our findings and test the reliability of the sensors with a larger sample size and under various driving conditions. Although our study demonstrated that physiological signals indicate increasing levels of cognitive load on drivers, we only examined three difficulty levels. Furthermore, our research concentrated solely on the cognitive load imposed by a secondary task, which represents only a portion of the external demands drivers face. Future studies should explore different types of secondary tasks, including other validated cognitive secondary tasks from earlier research, to determine how these physiological systems and vehicle dynamics respond. This expanded scope will provide a more comprehensive understanding of the factors affecting cognitive load in real-world driving scenarios.